\documentclass [11pt]{article}

\usepackage{amsmath, amsthm}

\usepackage{amssymb}
\usepackage{graphicx}
\usepackage{verbatim}

\usepackage[a4paper, top=1in, bottom=1.25in, left=1.25in, right=1.25in]{geometry}

\begin{document}

\title{\centering\Large{
An Incentive Analysis of some Bitcoin Fee Designs
                               }}
\date{}

\author{Andrew Chi-Chih Yao\thanks{Institute for Interdisciplinary Information Sciences, Tsinghua University, Beijing.}}
\maketitle{}

\begin{abstract}
In the Bitcoin system, miners are incentivized to join the system and validate transactions through fees paid by the users.  A simple ``pay your bid" auction has been employed to determine the transaction fees.  Recently, Lavi, Sattath and Zohar \cite{LSZ17} proposed an alternative fee design, called the \emph{monopolistic price} (MP) mechanism, aimed at improving the revenue for the miners. Although MP is not strictly incentive compatible (IC), they studied how close to IC the mechanism is for iid distributions, and conjectured that it is nearly IC asymptotically based on extensive simulations and some analysis. In this paper, we prove that the MP mechanism is nearly incentive compatible for any iid distribution as the number of users grows large. This holds true with respect to other attacks such as splitting bids. We also prove a conjecture in \cite{LSZ17} that MP dominates the RSOP auction in revenue (originally defined in Goldberg et al. \cite{GHKSW06} for digital goods). These results lend support to MP as a Bitcoin fee design candidate.  Additionally, we explore some possible intrinsic correlations between incentive compatibility and revenue in general.
\end{abstract}

\newpage

\section{Introduction}

\emph{Bitcoin}, and more broadly \emph{blockchain} systems, rely on the willingness of honest players to participate in the system. A good blockchain system should have simple, practical designs with suitable security guarantees against cheating. The \emph{incentive compatibility} (IC) concept which seeks to incentivize the participants to be truthful is playing an increasingly central role in the design of distributed financial systems.

Recently, Lavi, Sattath and Zohar \cite{LSZ17} started a study of the subject of \emph{Bitcoin Fee Market} design. In this market, there are two kinds of players: the \emph{users} who have transaction records that need to be certified and registered in the bitcoin system, and the \emph{miners} who create new blocks to include the transactions and get them certified.  Each user declares the maximal amount she is willing to pay for her transaction, and the miners use a mechanism to decide which transactions to include and how much fee to charge each user.  A primary focus of their study is the \emph{Monopolistic Price (MP)} mechanism, which is a natural and practical mechanism, although not IC in the strict sense (see Section 2 below). Their extensive simulations indicate that the mechanism does not deviate too much from being IC for most iii distributions, as the number of users $n$ grows large. An analysis was given for the special case of discrete distributions of finite size. They suggest that MP might be a good alternative to the ``pay your bid" auction, which is subject to low bids and revenue.  It is posed as a conjecture that the MP mechanism is nearly-IC for general iid distributions.

We will prove that MP is nearly-IC for any iid distribution as $n$ grows large, thus mathematically validating the strong simulation results obtained in \cite{LSZ17}. Note that the standard IC criterion (in auction market) only addresses one kind of attack, namely, the reporting of an untruthful bid value.  There are other possible attacks by Bitcoin users: for example, in the \emph{multiple strategic bids} (MSB) attack discussed in \cite{LSZ17}, a user could gain advantage by splitting his transaction into several transactions and bidding separately. This strategy could even enable a losing transaction to be included in the block. We will also prove the nearly-IC conjecture for MP with respect to the MSB attack.

We consider another mechanism, called the RSOP (Random Sampling Optimal Price) auction, which was first defined by Goldberg et al. \cite{GHKSW06} in the digital goods context and shown to be truthful. We prove that the RSOP revenue is always dominated by the MP revenue, as conjectured in \cite{LSZ17}.  In fact, the revenue difference can be arbitrarily large for some distributions, prompting us to look more deeply into possible correlation between IC and revenue in general (Theorem 6-8).

The contributions of the present paper are two-fold.  First, we prove the Monopolistic Price mechanism to be nearly-IC, confirming the previous strong experimental data. This holds true against the MSB attack as well. We also show MP to dominate the RSOP auction in revenue. These results lend support to MP as a Bitcoin fee design candidate.  Secondly, the methodology used in our proofs involves sophisticated mathematical analysis.  It demonstrates that theoretical computer science can provide powerful tools to complement system design for blockchains. Finally, we believe that the emerging area of incentive compatible blockchain design is an exciting research area with many intriguing problems to solve, for theorists and system designers alike.

\noindent\emph{Related Work:} The basic model for Bitcoin fee market introduced in \cite{LSZ17} in fact resembles the maximum revenue problem for \emph{Digital Goods} as considered by Goldberg et al. \cite{GHW01}\cite{GHKSW06}. The MP mechanism is similar to the optimal omniscient auction in \cite{GHKSW06}. However, The Bitcoin fee market differs from digital goods in its additional features: such as the auctioneer may delete or insert bids, or the users may split bids. This makes the Bitcoin fee design a rich and relevant new research subject for auction theory and mechanism design.  Among those work closely related to the current subject include Babaioff et al. \cite{BDOZ12}, Kroll et al. \cite{KDF13}, Carlsten et al. \cite{CKWN16},
Bonneau \cite{Bon16}, Huberman et al. \cite{HLM17}.  To formulate meaningful incentive models, there is much research work in Bitcoin which provides important ingredients for consideration. A more complete survey of related work can be found in \cite[Section 1.3]{LSZ17}.

\section{Review of the Models and Known Results}

We first review the Bitcoin fee model and several mechanisms defined and studied in \cite{LSZ17}.  A \emph{miner} acts as a monopolist who offers $n$ users to include their transactions in the miner's next block for a fee.  Each user $i$ has one transaction that needs this service and is willing to pay a fee up to some value $v_i$. The miner's problem is to design a mechanism to extract good revenue. The standard Bitcoin mechanism in use is a pay-your-bid system, where the miner simply takes the highest bids to fill the capacity of the block. This mechanism may not receive good revenue, since some bidders may not reveal the true value of the fees they are willing to pay. In view of this, some alternative mechanisms are proposed in \cite{LSZ17} and their security properties considered, which we will review below.

\subsection{The Monopolistic Price (MP) Mechanism}

Suppose user $i$ bids $v_i$ for $1\leq i\leq n$.  Sort $v_1, \cdots, v_n$ into a decreasing sequence $b_1\geq b_2\geq \cdots \geq b_n$.  Define the \emph{monopolistic price} as $R(v)=\max_{k\in \left[n\right]}k\cdot b_k$.  Denote by $k^\ast(v)$ the $k$ maximizing $R(v)$ (in case of ties, $k^\ast$ is taken to be the maximal one). The miner will include all users with the highest bids $b_1, b_2, \cdots  b_{k^\ast(v)}$ and simply charge each one of them the same fee $b_{k^\ast(v)}$. Call this the \emph{monopolistic price} $p^{mono}(v)=b_{k^\ast(v)}$. This mechanism gives the miner revenue $R(v)$, which is obviously the maximum revenue obtainable if all accepted bids must be charged a single price.

The above MP Mechanism is not truthful. \cite{LSZ17} analyzed how serious the non-truthfulness can be; we review their results below.

Consider any user $i$ and the vector of bids $v_{-i}=(v_1, \cdots, v_{i-1}, v_{i+1}, \cdots  v_n)$ of the other users. Let \\
$p^{honest}(v_i, v_{-i})=p^{mono}(v_i, v_{-i})$,  \ {and}\\
$p^{strategic}(v_{-i})=\min\{b_i\in R\, | p^{mono}(v_i, v_{-i})\leq v_i\}$.\\
To measure how much temptation there is for the users to \emph{shade} their bids, define the \emph{discount ratio} $\delta_i$ for user $i$:
\begin{align*}
\delta_i(v_i, v_{-i})=\begin{cases}
                         1-\frac{p^{strategic}(v_{-i})}{p^{honest}(v_i, \, v_{-i})}\  \ &\text{if}\  v_i \geq p^{strategic}(v_{-i}),\\
                         0                                                       \  \ &\text{otherwise}.
                        \end{cases}
\end{align*}
This ratio captures the gain a user can obtain by bidding strategically (instead of truthfully).

A major theme of \cite{LSZ17} is to investigate how large the discount ratio will typically be.  Assume all true values $v_i$ are drawn iid from some distribution $F$ on $[0, \infty)$.  Two measures are defined:  the worst-case measure and the average case one. For the former, let
\begin{align*}
              \delta_{max}(v)&=\max_i \delta_i(v_i, v_{-i}),\\
               \Delta_n^{max}&=E_{(v_1, \cdots, v_n)\sim F}[\delta_{max}(v)].\\
\end{align*}
\noindent{For the latter, let
\begin{align*}               \Delta_n^{average}=E_{(v_1, \cdots, v_n)\sim F}[\delta_1(v_1, v_{-1})].
\end{align*}
Clearly, for every $F$ and $n$, we have $\Delta_n^{max}\geq \Delta_n^{average}$, since $\delta_{max}(v)\geq \delta_i(v_i, v_{-i})$ for any $v$.

An analysis was given in \cite{LSZ17} for the special case of discrete distributions of finite size.

\vskip 8pt
\noindent{\textbf{Theorem A}}  \cite[Theorem 2.3]{LSZ17} \\
For any distribution $F$ with a finite support size, $\lim_{n\rightarrow\infty}\Delta_n^{max}(F)=0$. (This implies also
$\lim_{n\rightarrow\infty}\Delta_n^{average}(F)=0$ for such $F$.)

Based on extensive simulations done for a variety of distributions, the authors made the following conjecture that MP is nearly IC for general iid distributions as $n$ gets large.

\vskip 8pt
\noindent{\textbf{Nearly IC Conjecture for MP}} \cite[Conjecture 2.5]{LSZ17}

\noindent{1. For any distribution $F$, $\lim_{n\rightarrow\infty}\Delta_n^{average}(F)=0$.}  Specifically $\Delta_n^{average}(F)=O(\frac{1}{n})$.

\noindent{2. If $F$ has a bounded support, $\lim_{n\rightarrow\infty}\Delta_n^{max}(F)=0$.} Specifically $\Delta_n^{max}(F)=O(\frac{1}{n})$.

\noindent{3. There exists a distribution $F$ with an unbounded support such that  $\lim_{n\rightarrow\infty}\Delta_n^{max}(F)>0$.}

In the $O$-notation above, the constants may depend on $F$.  Part of the basis for their conjecture 3 is the Inverse distribution $F$ where $Pr_F\{X>x\}=\frac{1}{x}$ (for $x\in [1, \infty)$).  Experimentally, it appears that for this $F$,  $\lim_{n\rightarrow\infty}\Delta_n^{average}(F)=0$, while $\lim_{n\rightarrow\infty}\Delta_n^{max}(F)>0$.

The main purpose of our paper is to settle the Nearly IC Conjecture for MP in the positive.

\noindent\textbf{Multiple Strategic Bids (MSB) Attack}

It was shown in \cite{LSZ17} that a user could gain advantage by splitting his bid into several transactions with separate bids.  This strategy could even enable a losing transaction to be included in the block. Let
\begin{align*} p^{multi}(v_{-i})&=\min\left\{ u\cdot v^{(u)}_i |\, v^{(1)}_i\geq \cdots \geq v^{(u)}_i \geq p^{mono}(v_i^{(1)}, \cdots, v_i^{(u)},  v_{-i})\right\},\\
             \delta_i(v_i, v_{-i})&=\begin{cases} 1-\frac{p^{multi}(v_{-i})}{p^{honest}(v_i,\, v_{-i})}\ \ \ \text{if}\ \  v_i\geq p^{multi}(v_{-i}),\\
             0     \qquad\qquad\qquad\qquad    \text{otherwise.}
             \end{cases}
\end{align*}
It is conjectured that the Nearly-IC Conjecture holds even under this stronger attack.  We will show that this is indeed the case.

\subsection{The Random Sampling Optimal Price (RSOP) Auction}

We consider another mechanism, called the RSOP (Random Sampling Optimal Price) auction, first defined by Goldberg et al. \cite{GHKSW06} in the digital goods context.

\vskip 8pt
\noindent{{\emph{Definition.}}}  [RSOP auction] \\
Upon receiving $n$ bids $v=(v_1, \cdots, v_n)$, the auctioneer randomly partitions the bids into two disjoint sets $A$ and $B$, and computes the monopolistic price for each set: $P_A^{mono}$, $P_B^{mono}$ (with the monopolistic price for an empty set being set to $0$).  Finally, the set of winning bids is $A'\cup B'$, where $A'=\{i\in A:  v_i\geq P_B^{mono}\}$ and $B'=\{i\in B:  v_i\geq P_A^{mono}\}$.  The bidders in $A'$ each pays $P_B^{mono}$, and the bidders in $B'$ each pays $P_A^{mono}$.

Note the revenue obtained in this auction is
\begin{align*} RSOP(v)=|A'|\cdot P_B^{mono} + |B'|\cdot P_A^{mono}.
\end{align*}

\vskip 8pt
\noindent{\textbf{Theorem B}}  (Goldberg et al. \cite{GHKSW06}) \\
The RSOP auction is truthful.  For any $v=(v_1, \cdots, v_n)$ with $v_i\in [1, D]$ (where $D$ is a constant) for all $i$, we have
\begin{align*} \lim_{n\rightarrow\infty}\max_v\frac{R(v)}{RSOP(v)}=1.
\end{align*}

In \cite{LSZ17}, several variants of RSOP were examined and simulation carried out which led  to the following conjecture.

\vskip 8pt
\noindent{\textbf{Dominance Conjecture of MP over RSOP}} \cite[Conjecture 5.4]{LSZ17}\\
For any $v$ and all choices of $A$ and $B$, the RSOP revenue is at most the monopolistic price revenue.  That is, $RSOP(v)\leq R(v)$.

The MP Dominance Conjecture has relevance to the robustness of RSOP against adding false bids or deleting bids by the auctioneer (see discussions in \cite{LSZ17}). In the present paper we prove the MP Dominance Conjecture to be true.

\section{Main Results}

In this paper we settle the Nearly-IC Conjecture (even allowing for the MSB attack) and the MP Dominance Conjecture mentioned above: the former in Theorem 1-4, and the latter in Theorem 5. Additionally, we investigate the possible correlation between incentive compatibility and revenue.  In this regard, we demonstrate that distributions with unbounded support can exhibit different characteristics from the bounded ones, and these findings will be presented in Theorems 6-8.


\subsection{Nearly Incentive Compatibility of MP}

We prove that mechanism MP is nearly incentive compatible for large $n$ in Theorems 1-3.

\vskip 8pt
\noindent{\textbf{Theorem 1.}}  For any distribution $F$ on bounded support, $\lim_{n\rightarrow\infty}\Delta_n^{max}(F)=0$.

\vskip 8pt
\noindent{\textbf{Theorem 2.}}  For any distribution $F$, $\lim_{n\rightarrow\infty}\Delta_n^{average}(F)=0$.

\vskip 8pt
\noindent\emph{{Remark 1.}}  The proof in Theorem 1 can be refined to show that $\Delta_n^{max}(F)=O(\frac{1}{n^\beta})$ where $\beta>0$ is a constant independent of $F$, while the constant in the $O$-notation is $F$-dependent.  Similarly, Theorem 2 can be strengthened to $\Delta_n^{average}(F)=O(\frac{1}{n^\beta})$ when $F$ satisfies $\sup_x x(1-F(x))<\infty$.  The analysis follows the same outline as the proofs for Theorems 1, 2 above but the details are more complicated. They will be left for a later version of the paper.

Recall that the distribution Inverse is defined as $Pr_F\{X>x\}=\frac{1}{x}$ for $x\in [1, \infty)$.

\vskip 8pt
\noindent{\textbf{Theorem 3.}}  For $F=$Inverse,  $\lim_{n\rightarrow\infty}\Delta_n^{max}(F)>c$ for all $n$, where $c>0$ is some absolute constant.

\vskip 8pt
\noindent{\textbf{Theorem 4.}} With respect to the MSB attack, the Monopolistc Pricing Mechanism is nearly incentive compatible, i.e., Theorems 1-2 are still valid.

\vskip 8pt

\subsection{The Effect of IC on Revenue}

\noindent{\textbf{Theorem 5.}}  For any $v$, $RSOP(v)\leq R(v)$.

Theorem B of Goldberg et al. \cite{GHKSW06} says that, RSOP yields asymptotically the same revenue as the monopolistic price mechanism when the distribution $F$ has a bounded support $[1, D]$.  We point out that this is not always true when $F$ has infinite support.

\vskip 8pt
\noindent{\textbf{Theorem 6.}} For $F=$Inverse,
\begin{align*} &\lim_{n\rightarrow\infty}\frac{1}{n}E_{v_1, \cdots, v_n\sim F}[R(v)] = \infty,\  \text{while}\\
               &\lim_{n\rightarrow\infty}\frac{1}{n}E_{v_1, \cdots, v_n\sim F}[RSOP(v)] = 1.
\end{align*}

 Let $r_F=\sup_x x(1-F(x)$. For example, $r_F=1$ for $F=$ Inverse.  A key difference between RSOP and the monopolistic price mechanism is that, the former is incentive compatible (and thus cannot extract revenue more than $r_F n$), while the latter is not incentive compatible.

Suppose we are given $F=$ Inverse as value distribution. Theorem 6 above shows that Monopolistic Price can extract an unbounded revenue in this case. Is the property derived in Theorem 3 $\lim_{n\rightarrow\infty}\Delta_n^{max}(F)>c$ in fact a necessary condition for all such mechanisms? The following theorem shows that the answer is more complex.

For any mechanism $M$ and bid vector $v$, let $M(v)$ be the revenue collected.

\noindent{\textbf{Theorem 7.}}  Let $F=$ Inverse. \\
(a) There exists a mechanism $M$ such that $\lim_{n\rightarrow\infty}\Delta_n^{max}(F)=0$ and\\
 \begin{displaymath}\lim_{n\rightarrow\infty}\frac{1}{n}E_{v_1, \cdots, v_n\sim F}[M(v)] = \infty.\end{displaymath}
(b) Let $\eta <1$ be any fixed constant. Any mechanism $M$ such that $\delta_{max}(v)<\eta$ for all $v$ must satisfy\\
 \begin{displaymath}\lim_{n\rightarrow\infty}\frac{1}{n}E_{v_1, \cdots, v_n\sim F}[M(v)] <\infty.\end{displaymath}

Note that $\delta_{max}$ and $\Delta_n^{max}$ are not the only ways to quantify a mechanism's closeness to being incentive compatible. Two standard ways to define being $\epsilon$\emph{-close} to IC (or more generally, to Nash equilibrium) are \\
(1) Additively $\epsilon$-close:   $p(v_i, v_{-i}) - p_i^{mono}(v_{-i}) \leq \epsilon$, or  \\
(2) Multiplicatively $\epsilon$-close: $p(v_i, v_{-i}) - p_i^{mono}(v_{-i}) \leq \epsilon(v_i-p(v_i, v_{-i}))$.

We will show that, adopting the ``multiplicatively $\epsilon$-close" definition, one can obtain nearly IC mechanisms that derive infinite revenue under the distribution $F=$ Inverse. This is in contrast to the previous discount model defined in terms of $\delta_{max}$.

Let $M$ be any IC mechanism (such as RSOP).  For any bid vector $v$, let $p_i(v)$ be the fee paid by user $i$ (if $i$ is a winner), and $u_i(v)=v_i-p_i(v)$ be the \emph{utility}. For any $0\leq \epsilon <1$, let $M_\epsilon$ be the mechanism that uses the same allocation rule as $M$, with the fee modified to be $p'_i(v)=p_i(v) + \frac{\epsilon}{1+\epsilon} \, u_i(v)$.  The following theorem is easy to prove.

\vskip 8pt
\noindent{\textbf{Theorem 8.}}  \\
(a) $M_\epsilon$ is multiplicatively $\epsilon$-close to IC;\\
(b) \begin{displaymath}\lim_{n\rightarrow\infty}\frac{1}{n}E_{v\sim F}[M_\epsilon(v)]=
                       \lim_{n\rightarrow\infty}\frac{1}{n}E_{v\sim F}[M(v)] +
                       \frac{\epsilon}{1+\epsilon} \,\lim_{n\rightarrow\infty} \frac{1}{n}E_{v\sim F}(\sum _i u_i(v)).
\end{displaymath}
Theorem 8 implies that RSOP can be easily modified to become a  multiplicatively $\epsilon$-close-to-IC mechanism such that, like MP, its revenue is infinite under $F=$ Inverse.

We will prove Theorems 1-3, 5 in the following sections.  The proof for the Multiple Strategic Bids model of Theorem 4 is similar in essence to the basic model, and hence will be omitted. We also leave out the proofs of Theorem 6-8.

\section{An Overview of the Proof for Theorem 1}

Theorem 1 is the most difficult to prove.  In this section we give some intuition and an overview of the proof.

Let $F$ be a distribution over $[0, D]$. Let $v_1, \cdots, v_n$ be generated according to iid $F$, and denote by $b_1\geq b_2\geq \cdots \geq b_n$ the sorted list of the $v_i's$. By Claim A9 in \cite{LSZ17}, $\delta_{max}(b)=\delta_1(b_1, b_{-1})$, i.e. the maximum discount ratio is achieved by the user with the highest bid. This leads immediately to a necessary condition on $w$, the optimal strategic bid by the highest bidder, as we state below.

\noindent{\textbf{Lemma 1.}} [Optimal Strategic Bid (OSB) Condition]\\
Let $k^\ast=k^\ast(b)$.  If $\delta_{max}(b)\geq \eta$, where $0<\eta\leq 1$, then there exists $w\in[0, (1-\eta)b_{k^\ast}]$ such that $i_w\cdot w\geq k^\ast\cdot b_{k^\ast}-D$, where   $i_w$ is defined by $b_{i_w}\geq w > b_{i_w+1}$.

Lemma 1 states that, in order to have a sizable $\eta$, there has to exist a $w$ some distance away from $b_{k^\ast}$ such that $i_w\cdot w$ is only a constant $D$ smaller than the sampled maximum $R(b)=k^\ast\cdot b_{k^\ast}$.  We will prove Theorem 1 by showing that a random $b$ is stochastically unlikely to satisfy the OSB necessary condition.

As a start, we prove Theorem 1 when the distribution $F$ has a unique $\alpha_0>0$ where $A=\sup_\alpha\alpha(1-F(\alpha))$ is achieved.  The law of large number implies that, for large $n$, every $w\leq \alpha_0(1-\frac{1}{2}\eta)$ satisfies $i_w\cdot w<(A-\rho)n$ (where $\rho$ is some fixed constant) with overwhelming probability.  Coupled with the fact $b_{k^\ast}-\alpha_0=O(\frac{1}{\sqrt{n}})$ and $k^\ast\cdot b_{k^\ast}=A\cdot n+O(\sqrt{n})$ probabilistically, we see that the OSB condition in Lemma 1 cannot hold.  Hence we have shown Theorem 1 for the case when $\sup_\alpha\alpha(1-F(\alpha))$ has a maximum achieved at a unique point $\alpha=\alpha_0$.

The above argument does not apply when $\sup_\alpha\alpha(1-F(\alpha))$ achieves maximum value at multiple points. As an extreme case, consider the Inverse distribution modified as follows:
\begin{align*}
Inv^{(D)}(x)=\begin{cases}
                         1-\frac{1}{x}\  \ &\text{for}\  1\leq x<D,\\
                         1            \  \ &\text{for}\  x=D.
                        \end{cases}
\end{align*}
In this case, $x(1-Inv^{(D)}(x))=1$ for all $1\leq x<D$.

Hence the challenge is to prove that, even for extreme cases like the above, the OSB condition in Lemma 1 cannot be met except with vanishingly small probability. At the top level, we wish to show that, in any two subintervals $I, J\subseteq [0,D]$ separated by a non-negligible distance, the maximum values $A_I=\max\{j\cdot b_j\, | b_j\in I\}$ and $A_J=\max\{j\cdot b_j\, | b_j\in J\}$ cannot achieve perfect correlation to allow $|A_I-A_J|=O(1)$ except with negligible probability for large $n$.  This claim takes a non-trivial proof since intervals $A_I$ and $A_J$, being taken from the sorted version $b$ of $v$, are correlated to a certain degree.

Before proving Theorem 1, we first recast Lemma 1 in an new form which does not reference the quantity $w$. The main advantage of Lemma 1A is that, the condition now refers only to the quantities $b_1, \cdots, b_n$, thus making it easier to analyze how likely the condition can be satisfied stochastically.

\vskip 8pt
\noindent{\textbf{Lemma 1A.}} [Optimal Strategic Bid (OSB) Condition]\\
Let $k^\ast=k^\ast(b)$ and $0<\eta<1$. If $\delta_{max}(b)\geq \eta$, then there exists $b_j\in [0,\, b_{k^\ast}(1-\frac{1}{2}\eta)]$ such that
$j\cdot b_j\geq k^\ast\cdot b_{k^\ast}-\frac{2D^2}{\eta\cdot b_{k^\ast}}$.

\noindent\emph{Proof.}  Take the $w$ as specified in Lemma 1.  The following constraints are satisfied: write $i=i_w$ and $B= k^\ast\cdot b_{k^\ast}$, then
 \begin{align}  b_{k^\ast}- \eta\cdot b_{k^\ast} &\geq w, \\
                                              i\cdot w &\geq B-D.
 \end{align}
Let $\Delta_j=b_{k^\ast}- b_{j}$ for $j\in \{i, i+1\}$.  Let $0\leq \lambda <1$ such that $w=\lambda b_i + \lambda' b_{i+1}$ where $\lambda'=1-\lambda$.  It is easy to verify  from Eq. (1), (2) that
\begin{align} \lambda\Delta_i + \lambda' \Delta_{i+1} &\geq \eta\cdot b_{k^\ast},\\
               \lambda(i\cdot b_i) + \lambda' \left((i+1)b_{i+1}\right) &\geq B-D.
\end{align}
Note that Eq. (4) implies
\begin{align} \max\{ i\cdot b_i, \, (i+1)b_{i+1}\} \geq B-D.
\end{align}
We now prove Lemma 1A.  \\
\underline{Case 1}.  If $\Delta_i>\frac{1}{2}\eta\cdot b_{k^\ast}$, then choose $j\in \{i,\, i+1\}$ depending on which gives the larger $j\cdot b_j$. This $j$ satisfies Lemma 1A, as a consequence of Eq. (1) and (5). \\
\underline{Case 2}.  $\Delta_i \leq\frac{1}{2}\eta\cdot b_{k^\ast}$. Eq. (3) implies $\lambda'D \geq \frac{1}{2}\eta\cdot b_{k^\ast}$, and thus
\begin{align}   \lambda' \geq \frac{\eta\cdot b_{k^\ast}}{2D}.
\end{align}
From Eq. (4) and the fact $i\cdot b_i \leq B$, we have
\begin{align*}    \lambda' \left((i+1)b_{i+1}\right) \geq \lambda'B-D.
\end{align*}
Using Eq. (6), we obtin
\begin{align*}    (i+1)b_{i+1} \geq B-\frac{2D^2}{\eta\cdot b_{k^\ast}}.
\end{align*}
Taking $j=i+1$ satisfies Lemma 1A.               \qed

\section{Proof of Theorem 1}

For simplicity of presentation, we assume that $F$ has support $[1, D]$.  Without loss of generality, we can assume that $F(x)<F(D)=1$ for any $x<D$. Some additional arguments are needed for the general case $[0, D]$; we omit them here.

We will demonstrate that for a random $b$, the condition stated in Lemma 1A occurs with probability at most $O(\frac{(\log n)^2}{\sqrt n})$. That is, stochastically, the pair $(b_j, b_{k^\ast})$ with the stated property rarely exists. First some notation.  Let $H_F(x)=Pr_{z\sim F}\{z\geq x\}=1-F(x-)$.  To start the proof, we pick a point $D_1\in [1, D]$ with the following properties:

\noindent\textbf{P1:} $(D_1, D]$ is a forbidden zone for $b_{k^\ast}$.  Precisely, for a random $b$, the probability of $b_{k^\ast}\in (D_1, D]$ is $e^{-\Omega(n)}$.

\noindent\textbf{P2:} $Pr_{x\sim F}\{x\in [D_1, D]\} > 0$.

\vskip 8pt
\noindent{\textbf{Fact 1}}\ \ $D_1$ exists.

\noindent\emph{Proof.}  Let $\alpha_{max}$ be the maximal $\alpha$ achieving $\sup_\alpha\{\alpha\cdot H_F(\alpha)\}$.\\
\underline{Case 1}. If $\alpha_{max}=D$, then  $(D_1, D]$ is empty and $Pr_{x\sim F}\{x\in [D_1, D]\}=D\cdot H_F(D)\  > 0$.\\
\underline{Case 2}. If $\alpha_{max}<D$, then choose any $D_1\in (\alpha_{max}, D)$.  Choose $\Delta=\frac{1}{2}(D_1 - \alpha_{max})$.  Then for large $n$, the probability of $b_{k^\ast}\in [0, \, \alpha_{max}+\Delta]$ is $1-e^{-\Omega(n)}$, satisfying \textbf{P1}.  We also have $Pr_{x\sim F}\{x\in [D_1, D]\} \geq 1-F(D_1)>0$, thus satisfying \textbf{P2}.   \qed

Divide $[1, D_1]$ into disjoint intervals of length $\epsilon$, that is, write $[1, D_1]=\cup_{\ell=1}^m I_\ell$ where $I_\ell=[1+(\ell-1)\epsilon, \, 1+\ell \epsilon)$  for $1\leq \ell <m$, and $I_m=[1+(m-1)\epsilon, \, 1+m \epsilon]$.  Take a random $b$, which is the sorted list of iid $v_1, \cdots, v_n \sim F$.  Let $A_{\ell}^{max}$ denote the random variable $\max\{i\cdot b_i |\, b_i\in I_{\ell} \}$.  Let $A_{>\ell}^{max}$ be the random variable $\max\{i\cdot b_i\, |\, b_i\in I_{\ell+1}\cup\cdots \cup I_m \}$.  Let $W_\ell$ denote the event that
\begin{align*} A_{>(\ell+1)}^{max} - \frac{D^2}{\epsilon} \leq A_{\ell}^{max} \leq A_{>(\ell+1)}^{max}.
\end{align*}
Now note that, for $\delta_{max}(b) > 2\epsilon$, the OSB  condition for $(B_j, b_{k^\ast})$ in Lemma 1A can hold only if either 1) $b_j\in I_\ell$ and $b_{k^\ast}\in I_{\ell+2}\cup\cdots \cup I_m$ for some $\ell$, or 2) $b_{k^\ast}\in (D_1, D]$.

\vskip 8pt
\noindent{\textbf{Lemma 2.}}  \ \ $Pr\{\delta_{max}(b) > 2\epsilon\} \leq \sum_{\ell=1}^{m-2}Pr\{ W_\ell\} +e^{-\Omega(n)}$.

\noindent\emph{Proof.} Immediate from property \textbf{P1} and Lemma 1A.  \qed

The rest of this section is devoted to the proof of the following lemma, which indicates that $A_{\ell}^{max}$ and $A_{>(\ell+1)}^{max}$ are not correlated to be nearly identical.

\vskip 8pt
\noindent{\textbf{Lemma 3.}} [Weak Correlation Lemma]\ \ For each $1\leq \ell \leq m-2$, \  $Pr\{ W_\ell\} = O(\frac{(\log n)^2}{\sqrt n})$.

There are two cases to consider.\\
\noindent\underline{Case 1}.  $Pr_{x\sim F}\{x \in I_{\ell+1}\} > 0$;\\
\noindent\underline{Case 2}.  $Pr_{x\sim F}\{x \in I_{\ell+1}\} =0$.

We give the proof of Case 1. The proof for Case 2 uses the same general idea, and will only be sketched with details omitted here.  Assume we have Case 1.  Let $G$ be the distribution (normalized) when $F$ is restricted to the interval $L\equiv I_{\ell+1}\cup I_{\ell+2}\cup \cdots \cup I_m \cup [D_1, D]$.  Let
\begin{align*}  \rho_0&=Pr_{x\sim G}\{ x\in I_{\ell +1}\} > 0, \ \text{and}\\
                \rho_1&=Pr_{x\sim G}\{ x\in [D_1, D]\} > 0.
\end{align*}
Consider the generation of a random $b$ in the following alternative (but equivalent) way.

\noindent\underline{Phase 1}.  Generate a random integer $N$ so that
\begin{align*} Pr\{N=s\}= {n\choose s}q^s(1-q)^{n-s}
\end{align*}
for $0\leq s\leq n$, where $q=Pr_{x\sim F}\{x\in L\}$.

\noindent\underline{Phase 2}.  Generate $n-N$ iid random numbers $v_1, \cdots, v_{N-n} \sim F |_{I_1\cup\cdots\cup I_{\ell}}$, and sort them into decreasing order $b_{N+1}, b_{N+2},\cdots, b_n$.  (Note that $A^{max}_{\ell}$ is already determined after the completion of Phase 2.)

\noindent\underline{Phase 3}.  Generate $v_1, v_2, \cdots, v_{N} \sim G$ one number at a time.  After step $t$, we sort the numbers $v_1, v_2, \cdots, v_t$ into decreasing order $b^{(t)}_1 \geq b^{(t)}_2 \geq \cdots\geq b^{(t)}_t$.  Let
\begin{align*} A^{(t)}= \max\{i\cdot b_i^{(t)} \ | \ b_i^{(t)}\in  I_{\ell+2}\cup I_{\ell+3}\cup\cdots \cup I_m\}.
\end{align*}
At time $t=N$, $A^{(N)}$ is exactly the same random variable as $A^{max}_{>(\ell+1)}$.

We prove two facts below, from which Lemma 3 follows immediately.

\vskip 8pt
\noindent{\textbf{Fact 2}} \ \  In Phase 1, $N\geq \frac{1}{2}\, q \,n $ with probability $1 - e^{-\Omega(n)}$.

\noindent\emph{Proof.} Chernoff's bound.  \qed

After Phase 1 and 2, we have $N$ and $K=A^{max}_{{\ell}}$ decided.  To prove Lemma 3 we only need to show that, in Phase 3, there is enough randomness so that $A^{(N)}$ is unlikely to have a value within an additive constant $\frac{D^2}{\epsilon}$ to $K$.

Let us examine the evolution of $A^{(t)}$ as a random process of infinite length.  The random sequence $A^{(t)}$ satisfies $A^{(0)} =0$, and
\begin{align*}
A^{(t)}\begin{cases}
                      =A^{(t-1)}         &\qquad \text{with probability} \ \rho_0,   \\
                     \geq   A^{(t-1)}+1 &\qquad \text{with probability} \ \rho_1,  \\
                     \geq   A^{(t-1)}  &\qquad \text{otherwise},
                    \end{cases}
 \end{align*}
for $t\geq 1$.

\vskip 8pt
\noindent{\textbf{Fact 3}}\ \  At $t=N$, we have
\begin{displaymath} Pr\{|A^{(N)}-K|<\frac{D^2}{\epsilon}\}= O(\frac{(\log N)^2}{\sqrt N})=O(\frac{(\log n)^2}{\sqrt n}).
\end{displaymath}

\noindent\emph{Proof.} (Sketch).
Let $s$ be the total number of times in the above process when either the second or third selection is made by $A^{(t)}$, for $1\leq t\leq N$. Let $A^{(t_1)}, A^{(t_2)}, \cdots, A^{(t_s)}$ be the projected sequence. Note $E(s)=(1-\rho_0)N$, $Var(s)=\Theta(N)$, and in fact $Pr\{s=u\}=O(\frac{1}{\sqrt N})$ for any $u$.  Relabel $A^{(t_i)}$ as $B^{(i)}$, and consider the random sequence $B^{(1)}, B^{(2)}, \cdots$.  Let $B^{(i)}, B^{(i+1)}, \cdots, B^{(i')}$ be the portion of the sequence in the range $[K-\frac{D^2}{\epsilon}, \, K+\frac{D^2}{\epsilon}]$.  It is easy to verify that, for the random sequence $B^{(1)}, B^{(2)}, \cdots,$
\begin{align} Pr\{i'-i+1 \geq \frac{D^2}{\epsilon}(\log N)^2\}=O(N^{-\log N}).
\end{align}
 To see this, note that there is at least a constant $\phi=\frac{\rho_1}{1-\rho_0}$ probability to increase the next $B^{(j)}$ value by 1 (or more).  Thus, to increase the value by $\frac{D^2}{\epsilon}$, it takes only $\frac{D^2}{\epsilon}\frac{1}{\phi}$ steps on average, rarely requiring a $(\log N)^2$ factor more steps.

 As $Pr\{s=u\}=O(\frac{1}{\sqrt N})$ for any $u$, we conclude that $Pr\{B^{(s)}\in [K-\frac{D^2}{\epsilon}, \, K+\frac{D^2}{\epsilon}]\} \leq O(\frac{(\log n)^2}{\sqrt n})$. This completes the proof of Case 1.

 In Case 2, $Pr_{x\sim F}\{x\in I_{\ell+1}\}=0$.  We have lost the source of randomness critical for the above argument since $\rho_0=0$. Yet the source of randomness can be obtained by splitting $I_\ell$ into $I'_\ell\cup I^{''}_\ell$ suitably, so that $I^{''}_\ell$ does not contain any $b_j$ with $j\cdot b_j$ anywhere close to the level of $A^{max}_{>(\ell+1)}$. (This will rely critically on the fact $Pr_{x\sim F}\{x\in I_{\ell+1}\}=0$.)  We omit here the details of implementing this plan, as well as the necessary handling of discontinuities in distribution $F$.  This finishes the proof of Lemma 3 and thus Theorem 1. \qed.

\section{Proof of Theorem 2}

For any $1>\epsilon>0$, we show that
\begin{align} \Delta^{average}_n (F)\leq \epsilon
\end{align}
for all sufficiently large $n$.  First pick $D>0$ such that $F(D)>1-{\epsilon}/{3}$.  Take $n$ iid $v_i\sim F$ and let $q_{n,m}$ be the probability of exactly $m$ of the $v_i's$ falling into $[0, D]$.  Let $ \epsilon'=\epsilon/2$.  By the law of large numbers, there exists $N_1$ such that for all $n\geq N_1$,
\begin{align} \sum_{m\leq(1-\epsilon')n} q_{n,m} < \epsilon/6.
\end{align}
Consider the distribution $G$, obtained from restricting $F$ to $[0, D]$.  By Theorem 1, there exists $N_2>0$ such that
\begin{align} \Delta^{average}_m(G) < \epsilon/3
\end{align}
for all $m>N_2$. We are now set for proving Theorem 2.  By definition of $\Delta^{average}_n $, we have
\begin{align*} \Delta^{average}_n(F) \leq \sum_{m>(1-\epsilon')n} q_{n,m} \left[\frac{m}{n}\Delta^{average}_m(G) +  \frac{n-m}{n}\right] +\sum_{m\leq(1-\epsilon')n} q_{n,m}.
\end{align*}
Using Eq. (9)-(10), we obtain for all $n>\max\{N_1, N_2\}$,
\begin{align*}\Delta^{average}_n(F) \leq (\frac{\epsilon}{3}+\epsilon') + \frac{\epsilon}{6}=\epsilon.
\end{align*}
This proves Eq. (8) and Theorem 2.  \qed

\section{Proof of Theorem 3}

Consider $n$ iid random variables $v_1, \cdots, v_n$ distributed according to the Inverse distribution $Inv$, and let $b_1 \geq \cdots \geq b_n$ be their sorted sequence.  Let $\lambda=40$, $\lambda'=1$, and let $T_n$ be the event $(b_1>\lambda n)\wedge (b_2 < \lambda' n)$. Let $V_n$ be the event $(\max_{2\leq i \leq n} i\cdot b_i \leq \lambda n)$.  Theorem 3 is an immediate consequence of the following two lemmas.

\vskip 8pt
\noindent{\textbf{Lemma 4.}}\  If $v$ satisfies event $T_n \wedge V_n$, then $\delta_{max}(v) \geq 1 - \frac{\lambda'}{\lambda}$.

\vskip 8pt
\noindent{\textbf{Lemma 5.}} \begin{displaymath}Pr\{T_n \wedge V_n\} \geq \frac{1}{\lambda} e^{-\frac{2}{\lambda'}} - 4\, e ^{-\frac{\lambda}{2}}.
\end{displaymath}.

Lemma 4 and 5 imply
\begin{align*}\Delta^{max}_n(Inverse) \geq c,
\end{align*}
where \begin{displaymath} c= (1 - \frac{\lambda'}{\lambda})\cdot\left(\frac{1}{\lambda} e^{-\frac{2}{\lambda'}} - 4\, e ^{-\frac{\lambda}{2}}\right) >0.
\end{displaymath}
This proves Theorem 3.

To prove Lemma 4, note that when $T_n \wedge V_n$ occurs, the highest bidder has a monopolistic price $b_1$ and a strategic price less than $\lambda n$.  Thus for the highest bidder $i$, we have $\delta_i(v) \geq 1 - \frac{\lambda n}{b_1} \geq 1 - \frac{\lambda'}{\lambda}$.  This proves Lemma 4.

Lemma 5 follows from the following facts:

\vskip 8pt
\noindent{\textbf{Fact 4}} \ \  $Pr\{T_n\} \geq \frac{1}{\lambda} e^{-\frac{2}{\lambda'}}$.

\noindent\emph{Proof.}
\begin{align*}Pr\{T_n\} &= {n\choose 1}\frac{1}{\lambda n} \left(1 -\frac{1}{\lambda' n} \right)^ {n-1}\\
&\geq \frac{1}{\lambda} e^{-\frac{2}{\lambda'}}.
\end{align*}
\qed

\vskip 8pt
\noindent{\textbf{Fact 5}} \ \ $Pr\{V_n|T_n\} \leq 2\, e^{-\frac{\lambda}{2}}$.

 \noindent\emph{Proof.}  Let  $y_1, \cdots, y_n$ be iid distributed according to $G$, where for $t \in[1, n]$,
\begin{align*}Pr_{z\sim G}\{z>t\} = \frac{1}{1-\frac{1}{n}}(\frac{1}{t} - \frac{1}{n}).
\end{align*}
Define $Y_n^{max}=\max_{1\leq i\leq m}\{i\cdot b_i\}$, where $b_1 \geq \cdots \geq b_n$ is the sorted sequence of $y_1, \cdots, y_n$.  Clearly,
\begin{align*} Pr\{V_n|T_n\} \leq Pr\{Y_n^{max} \geq \lambda n\}.
\end{align*}
To prove Fact 5, it suffices to prove:
\begin{align} Pr\{Y_n^{max} \geq \lambda n\} \leq 2\, e^{-\frac{\lambda}{2}}.
\end{align}
For any $t\geq 1$, let $M_t$ be the number of $y_i$'s satisfying $y_i\geq t$, and $B_t$ be the event that $t\cdot M_t\geq \frac{\lambda}{2} n$.  Let $t_k=\frac{1}{2^k} n$ for $1\leq k \leq \lfloor \log_2 n\rfloor$.  As the event $Y_n^{max} \geq \lambda n$ implies $\vee_{1\leq k \leq \lfloor \log_2 n\rfloor} B_{t_k}$, we have
\begin{align} Pr\{Y_n^{max}\geq  \lambda n\}\leq \sum_{i=1}^{\lfloor\log_2 n\rfloor} Pr \{B_{t_k}\}.
\end{align}
Observe that $E(M_t)=(1-\frac{1}{n})\frac{n}{t}$.  Using Chernoff's bound, we have
\begin{align} Pr\{B_t\} \leq  e^{-\frac{\lambda}{2}\frac{n}{t}}.
\end{align}
Equation (11) follows from (12) and (13) immediately.  This completes the proof of Fact 5 and Theorem 3.  \qed

\section{Proof of Theorem 5}

Without loss of generality, we assume that $A$, $B$ are non-empty and that $p_A^{mono} \leq p_B^{mono}$.  Let $A$ consist of $y_1\geq \cdots \geq y_m$ and $B$ consist of $z_1\geq \cdots \geq z_\ell$, with $y_s=p_A^{mono}$ and $z_t=p_B^{mono}$.  Let $A'=\{y_1, y_2, \cdots, y_{s'}\}$ and $B'=\{z_1, z_2, \cdots, z_{t'}\}$ be the  winners from $A$ and $B$ respectively.  By definition of RSOP,
\begin{align*} y_s &\leq z_{t'}, \\
              t'z_{t'}  &\leq t\,z_{t}.
\end{align*}
It follows that
\begin{align*} RSOP(v)&=t'y_{s} + s'z_{t} \\
                      &\leq t'z_{t'}  + s'z_{t}\\
                       &\leq t\,z_{t}  + s'z_{t}.
\end{align*}
Now by definition $R(v)\geq (t+s')z_{t}$.  Theorem 5 follows.

\end{document}